\documentclass[]{aa}
\usepackage[dvips]{graphicx}
\usepackage{txfonts}
\usepackage{natbib}
\bibpunct{(}{)}{;}{a}{}{,}

\begin{document}

\title{RXTE determination of the intermediate polar status of XSS~J00564+4548, IGR~J17195--4100, and XSS~J12270--4859}

\author{O.W. Butters\inst{1}
\and
  A.J. Norton\inst{1}
\and
  P. Hakala\inst{2}
\and
  K. Mukai\inst{3}
\and
  E.J. Barlow\inst{1}
}

\institute{
  Department of Physics and Astronomy, The Open University, Walton
  Hall, Milton Keynes MK7 6AA, UK\\
  \email{o.w.butters@open.ac.uk}
\and
  Tuorla Observatory, University of Turku, FIN-21500 Piikki\"{o}, Finland
\and
  CRESST and X-ray Astrophysics Laboratory NASA/GSFC, Greenbelt,
        MD 20771, USA
}

\authorrunning{Butters et al.}
\titlerunning{RXTE analysis of IP candidates.}

\date{Accepted 2008 ???;
      Received  2008 ???;
      in original form 2008 ???}

\abstract
{}
{We determine the nature of the intermediate polar candidates \object{XSS~J00564+4548},
  \object{IGR~J17195--4100}, and \object{XSS~J12270--4859}.}
{Pointed {\em RXTE\em} observations searched for intermediate
    polar characteristics in these candidate systems.}
{\object{XSS~J00564+4548} exhibits a period of 465.68$\pm$0.07~s, which we
  interpret as the spin period, an energy dependent modulation depth,
  and a spectrum that is fit by a 22~keV
  photoelectrically absorbed bremsstrahlung with an iron line profile.
  \object{IGR~J17195--4100} shows several candidate
  periodicities and a spectrum that is fit by a power law with an iron line.
  \object{XSS~J12270--4859}
  exhibits a candidate spin period of 859.57$\pm$0.64~s and a spectrum that is fit
  by a power law with no evidence of an iron line.}
{\object{XSS~J00564+4548} is confirmed to be an intermediate
  polar. \object{IGR~J17195--4100} and \object{XSS~J12270--4859} both show some properties
  of intermediate polars, but cannot be confirmed as definite members
  of the class here.}


\maketitle

\section{Introduction to magnetic cataclysmic variables}

Intermediate polars (IPs) belong to the class of systems known as
cataclysmic variables (CVs). They occupy the phase space, in terms
of magnetic field strength, between the polars and the
non-magnetic CVs. This intermediate strength magnetic field
alters the accretion flow from the main
sequence donor star to the white dwarf (WD). Eventually, most of the
accreting material is channelled to accretion
curtains above the WD magnetic poles. The
temperature and density of this region causes the emission of
bremsstrahlung radiation, which varies at the spin period of
the WD. It is this variation that most consider to be the defining
characteristics of IPs. For a review of IPs see e.g.~\citet{warner95}.
There are at least 30 confirmed
IPs\footnote{asd.gsfc.nasa.gov/Koji.Mukai/iphome/iphome.html as at
30/04/08}. \citet{ramsay08}, however,
have recently pointed out that the commonly used criteria to certify
CVs as IPs may be too restrictive. It is possible that many of the
84 candidates\footnotemark[1] are indeed IPs, and if classes such as SW~Sex systems
are in fact IPs then the true number may be several hundred.

In recent years there have been 16 IPs found to emit in the
hard X-ray/soft gamma-ray part of the spectrum, with the
{\em INTEGRAL}/IBIS survey \citep{barlow06,bird07}.
With this in mind we have embarked on a campaign to observe some hard
X-ray sources and determine their credentials as potential
IPs. In the first paper in this campaign, \object{SWIFT~J0732.5--1331}
was confirmed as an IP \citep{butters07}. Here the
results of pointed {\em RXTE} observations of \object{XSS~J00564+4548}
(hereafter J0056), \object{IGR~J17195--4100} (hereafter J1719)
and \object{XSS~J12270--4859} (hereafter J1227) are presented.

\section{Previous observations}

J0056 was associated with the {\em ROSAT} source
  \object{1RXS~J005528.0+461143}, and catalogued as an unidentified
  object in the {\em RXTE} all sky survey \citep{revnivtsev04}. It was
  found to have a count rate of 0.71$\pm$0.04~ct~s$^{-1}$~PCU$^{-1}$
  in the 3--8~keV energy band and a photon index of 1.77$\pm$0.23. Analysis by
\citet{bikmaev06} using {\em SWIFT}/XRT archive data revealed two X-ray
sources in the {\em ROSAT} error circle. One source was present at low energy, which
they presumed to be a chromospherically active star. The
other source showed a typical spectrum of a CV, with
an emission feature close to 6.7~keV. \citet{bikmaev06} also carried
out optical observations with the 1.5~m Russian-Turkish Telescope.
Their photometric data indicated a period of approximately 480~s to be present.

J1719 was detected as an {\em INTEGRAL }object by \citet{bird04},
\citet{pandey06} found radio galaxies coincident with its error
circle and suggested it was extragalactic. \citet{tomsick06}
confirmed a tentative association of J1719 with the softer X-ray target
\object{1RXS~J171935.6--410054} using pointed {\em Chandra} data. They
also reported variability of J1719 in the 0.3--10~keV band and a
flux of $2.5^{+0.9}_{-0.4}\times10^{-11}$~ergs~cm$^{-2}$~s$^{-1}$. In
calculating this flux they used a power law model and a galactic
column density of 0.77$\times10^{22}$~cm$^{-2}$ (derived from
\citet{dickey90}). \citet{tomsick06} also reported the spectral
properties of J1719 using public {\em INTEGRAL} data, finding
a flux of $1.9\times10^{-11}$~ergs~cm$^{-2}$~s$^{-1}$ in the
20--50~keV energy band. \citet{masetti06}
classified J1719 as a CV based upon its optical
spectrum, they also speculated that it may be an IP.

J1227 was found in the {\em RXTE} all sky survey
\citep{revnivtsev04}. It was classified as a CV and suggested to be
an IP by \citet{masetti06}, using optical spectroscopy. \citet{bird07}
later found J1227 to be an {\em INTEGRAL} source.

\section{Observations and data reduction}
Data were obtained from the {\em RXTE} satellite \citep{bradt93} with the PCA
instrument.

\begin{table}
  \centering
  \caption{Observing log.}
  \label{observing_log}
  \centering
  \begin{tabular}{ccccc}
    \hline\hline
    Target   & Start time          & End time         & Time        & Good\\
       & (UTC)         & (UTC)        & on target   & time$^a$ \\
       &               &              & (s)     & (s) \\
    \hline
    J0056    & 05:27 20/12/07        & 00:31 22/12/07     & 84\,672 & 37\,800 \\
    J1719    & 18:45 07/01/08        & 11:16 09/01/08     & 69\,636 & 35\,936 \\
    J1227    & 16:13 28/11/07        & 16:20 29/11/07     & 58\,183 & 26\,814 \\
    \hline
    \multicolumn{5}{l}{(a) - Good time is defined as the time that met
    our selection criteria.}\\
  \end{tabular}
\end{table}

In each case initial data reduction was
done with the standard {\sc ftools}, and the flux was normalised according
to the number of correctly functioning PCUs. For the
lightcurve analysis PCUs 2, 3, and 4 were used; whilst for the spectral
analysis only PCU 2 was used. Only the top layer of
each PCU was included in the measurements and the time resolution of
the data was 16~s. Background subtracted light curves were constructed in four
energy bands: 2--4~keV, 4--6~keV, 6--10~keV and 10--20~keV, as well as
a combined 2--10~keV band for maximum signal-to-noise.

In the presence of white noise in the data, the power values in
the power spectrum are expected to follow an exponential
distribution. However, any correlated noise e.g. red noise, will
mean the distribution becomes frequency dependent. This makes estimating the
significance limits in the power spectra non-trivial. As accreting
systems usually show flickering in their lightcurves, it is
feasible to believe that there may be a significant red noise
component in the data. In order to take this into account
in the analysis, the technique introduced in
\citet{hakala04} was used. The data were equally spaced (apart from the
large gaps in between different orbits), so the red noise
component was modelled by fitting a second order autoregressive
process model to the lightcurves. This model was then used to generate
50\,000 synthetic lightcurves with similar red and white noise
properties, as well as observing window, to the original datasets.
The 95.2\%, 99.72\% and 99.954\% (2,
3 and 4$\sigma$ respectively) significance limits (as a function
of frequency) were then calculated.

To estimate the error on the measured periods we folded the raw
  data at the period found from the period analysis. We then fitted a
  curve to this folded data. This curve (repeated over the whole data set) was
  then subtracted from the raw data leaving residual values. These were then shuffled and added to
  the fitted curve, yielding a new synthetic raw data set. This
  synthetic data was then analysed as before. This whole process was
  repeated $\sim200$ times and the resulting periods were then used to
  calculate a standard deviation of periods, which was then used as
  the error estimate.

A mean X-ray spectrum was also extracted for each source, and two
spectral models applied to find the best fit, using the {\sc
xspec} package. The models considered were a photoelectrically
absorbed single temperature bremsstrahlung with a Gaussian at the
iron line emission energy (model A), and a photoelectrically absorbed power
law with a similar Gaussian (model B).

\subsection{XSS~J00564+4548}
J0056 was observed over two consecutive days (see Table~\ref{observing_log}).
The total good time on target (37\,800~s) comprised fourteen approximately
equal segments of one satellite orbit each. In the 2--10~keV energy band
the raw count rate varied between 3.9 and 9.1~ct~s$^{-1}$~PCU$^{-1}$. The
background count rate, generated from the calibration files, varied
between 2.9 and  4.1~ct~s$^{-1}$~PCU$^{-1}$.

\begin{figure}
  \resizebox{0.8\hsize}{!}{\rotatebox{90}{\includegraphics{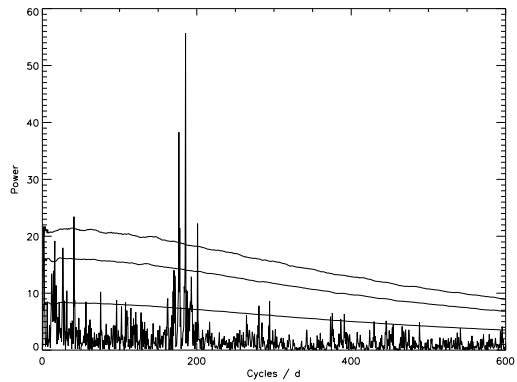}}}
  \caption{2~--~10~keV periodogram of J0056. Three significance
  levels, 95.2, 99.7 and 99.954\% (2, 3 and 4$\sigma$ respectively), are superimposed.}
  \label{J00564_red_noise}
\end{figure}

\begin{figure}
  \resizebox{0.8\hsize}{!}{\includegraphics{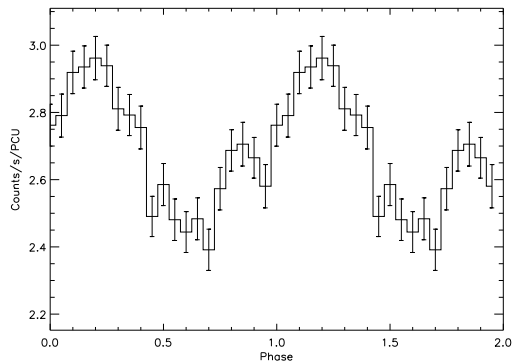}}
  \caption{2--10~keV lightcurve of J0056 folded at the 465.68~s pulse period
  with an arbitrary zero point. Two cycles are shown for clarity.}
  \label{J00564_folded}
\end{figure}

A significant ($>4\sigma$) peak was present in the periodogram at
$\sim$185~cycles day$^{-1}$ in the 2--10~keV energy band
(see Fig.~\ref{J00564_red_noise}). Analysis of the peak gave a pulsation period of
465.68$\pm$0.07~s. The data were then folded in each energy band at this
period, Fig.~\ref{J00564_folded} shows the result of the 2--10~keV
energy band. In each energy band a sinusoid was fitted to the folded
data to estimate the modulation depth of the variation
(see Table~\ref{modulation_depths}). There is a clear decreasing trend in
the modulation depth with increasing energy.
Clustered around the 185 cycles day$^{-1}$ peak were a series of smaller peaks,
spaced apart by $\sim8$ cycles day$^{-1}$, the largest of which was at
489.0$\pm$0.7~s. There was also one other peak detected at above the
4$\sigma$ level at $\sim41$ cycles~day$^{-1}$ (2\,109~s).

\begin{table*}
  \centering
  \caption{Modulation depths of the pulse profile in different energy bands.
  Modulation depth is defined here as the semi-amplitude of a fitted sinusoid
  compared to the fitted mean. }
  \label{modulation_depths}
  \centering
  \begin{tabular}{l@{\hspace{1em}}r@{--}l l@{\hspace{2em}}r@{$\pm$}l ccccc}
    \hline\hline
    \multicolumn{4}{c}{\,}      & \multicolumn{3}{c}{J0056$^{\rm a}$}         & \multicolumn{2}{c}{J1719$^{\rm b}$}         & \multicolumn{2}{c}{J1227$^{\rm c}$}\\
    \multicolumn{3}{c}{Energy band} & \multicolumn{3}{c}{Modulation depth} & Fitted mean              & Modulation depth & Fitted mean              & Modulation depth & Fitted mean\\
    \multicolumn{3}{c}{(keV)}       & \multicolumn{3}{c}{(\%)}             & (ct~s$^{-1}$~PCU$^{-1}$) & (\%)             & (ct~s$^{-1}$~PCU$^{-1}$) & (\%)             & (ct~s$^{-1}$~PCU$^{-1}$)\\
    \hline
    &2&10            && 8&1          & 2.69         & 5$\pm$1          & 4.22             &   26$\pm$2           & 1.29  \\
    &2&4             && 14&2         & 0.59         & 4$\pm$1          & 1.01             &   27$\pm$3           & 0.37  \\
    &4&6             && 9&1          & 0.93         & 4$\pm$1          & 1.45             &   25$\pm$3           & 0.45  \\
    &6&10            && 5&1          & 1.17         & 6$\pm$1          & 1.75             &   27$\pm$3           & 0.47  \\
    &10&20           && 8&3          & 0.60         & 5$\pm$2          & 0.72             &   28$\pm$7           & 0.22  \\
    \hline
    \multicolumn{11}{l}{(a)~-~Folded at the 465.68~s period, (b)~-~Folded at the possible period of 1\,842.4~s, (c)~-~Folded at the 859.57~s period.}\\
  \end{tabular}
\end{table*}

\begin{figure}
  \resizebox{0.8\hsize}{!}{\rotatebox{-90}{\includegraphics{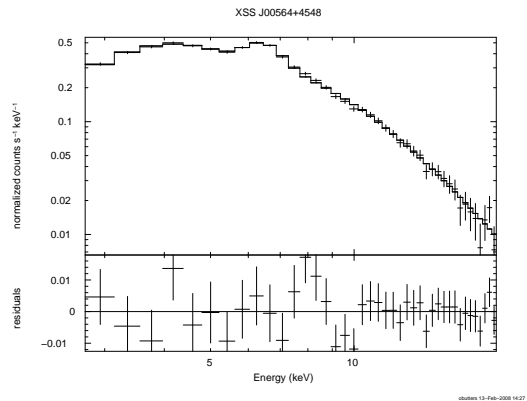}}}
  \caption{2.5--20 keV mean spectrum of J0056 fitted with a photoelectrically
  absorbed bremsstrahlung plus iron line profile.}
  \label{J00564_spectrum}
\end{figure}

The best spectral fit was a
simple photoelectrically absorbed bremsstrahlung model with a Gaussian
added. This fit had the parameters $kT=22\pm2$~keV, $n_{\rm
  H}=(0.6\pm0.4)~\times10^{22}$~cm$^{-2}$ and a Gaussian at
6.5$\pm$0.1~keV with a width of 0.3$\pm$0.1~keV, which was
interpreted as a iron feature ($\chi_{\rm reduced}^2=0.8$), as shown in
Fig.~\ref{J00564_spectrum} and summarised in
Table~\ref{spectral_fits}. The table also shows the Galactic column
density to the object as derived from the HEASARC $n_{\rm H}$
tool\footnote{http://heasarc.nasa.gov/cgi-bin/Tools/w3nh/w3nh.pl}.

\begin{table*}
  \centering
  \caption{Spectral fits.}
  \label{spectral_fits}
  \centering
  \begin{tabular}{ccccccccccc}
    \hline\hline
    Target  & $n_{\rm H}$(Galactic) & Model &
    $n_{\rm H}$           & $kT$     & $\Gamma$    & Fe       &
    $\sigma_{\rm Fe}$ & EW       & $\chi^2_{\rm reduced}$ & Flux (2--10keV)\\
      & ($\times10^{22}$~cm$^{-2}$)     &           & ($\times10^{22}$~cm$^{-2}$) & (keV)    &             & (keV)        & (keV)                  & (keV) &                        & ($\times10^{-11}$~ergs~cm$^{-2}$~s$^{-1}$)\\
    \hline
    J0056     & 0.1               & A           & 0.6$\pm$0.3           & 22$\pm$3 & --          & 6.5$\pm$0.1  & 0.3$\pm$0.1       & 0.8      & 0.8           & 2.8 \\
    J0056     & 0.1               & B           & 1.9$\pm$0.7           & --       & 1.7$\pm$0.1 & 6.5$\pm$0.1  & 0.3$\pm$0.1       & 0.9      & 1.1                & 2.9 \\
    J1719     & 0.7               & A           & 0.7$\pm$0.1           & 17$\pm$1 & --          & 6.5$\pm$0.1  & 0.1$\pm$0.1       & 0.5      & 3.0           & 4.4 \\
    J1719     & 0.7               & B           & 0.7$^{\rm a}$         & --       & 1.8$\pm$0.1 & 6.5$\pm$0.1  & 0.3$\pm$0.1       & 0.7      & 1.1                & 4.7 \\
    J1227     & 0.1               & A           & 0.1$^{\rm a}$         & 14$\pm$1 & --          & 6.5$^{\rm b}$& 0.1$^{\rm b}$     & $<$0.08 & 1.3                & 1.5 \\
    J1227     & 0.1               & B           & 0.1$^{\rm a}$         & --         & 1.8$\pm$0.1 & 6.5$^{\rm b}$& 0.1$^{\rm b}$       & $<$0.17 & 0.8                & 1.5 \\
    \hline
    \multicolumn{11}{l}{(a) - Pegged to a lower limit of this value to reflect the Galactic column density, (b) - No error as this value was imposed. See notes in the text.}\\
  \end{tabular}
\end{table*}

\subsection{IGR~J17195--4100}
Data were taken over two consecutive days (see
Table~\ref{observing_log}). The total good time on target (35\,936~s)
was split over twelve approximately equal segments. The raw target flux
varied from 5.4--11.3~ct~s$^{-1}$~PCU$^{-1}$ and the generated
background varied from 2.8--3.9~ct~s$^{-1}$~PCU$^{-1}$.

\begin{figure}
  \resizebox{0.8\hsize}{!}{\rotatebox{90}{\includegraphics{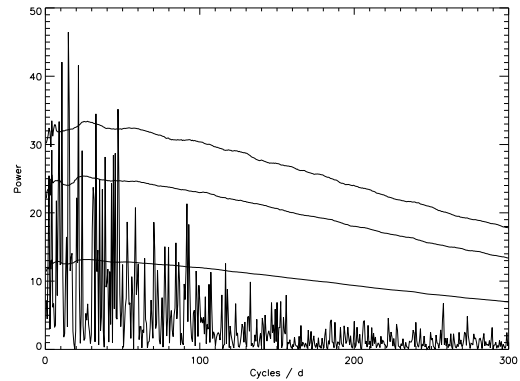}}}
  \caption{2--10~keV periodogram of J1719. Three significance
  levels, 95.2, 99.7 and 99.954\% (2, 3 and 4$\sigma$ respectively), are superimposed.}
  \label{J17195_red_noise}
\end{figure}

\begin{figure}
  \resizebox{0.8\hsize}{!}{\includegraphics{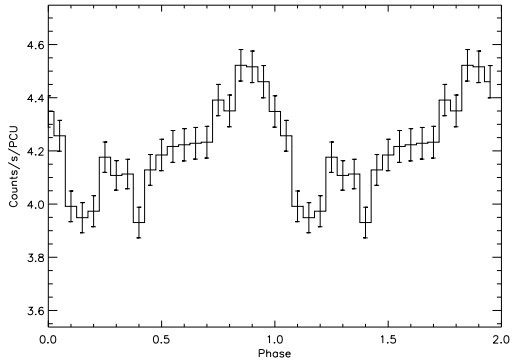}}
  \caption{2--10~keV folded lightcurve of J1719. Folded at 1\,842.4~s with an arbitrary zero point. Two periods are shown for clarity.}
  \label{J17195_2_10kev_folded}
\end{figure}

The periodogram of J1719 had six potential periods that were over
4$\sigma$, see Fig.~\ref{J17195_red_noise}. To discount any
  artifacts arising from the windowing of the raw data we also used
  the {\sc clean} algorithm of \citet{lehto97}. This was a
  necessary step as the raw data was rather fragmented. This iteratively 
  deconvolved the window function from any signals present in the lightcurve itself. The four peaks
between 8 and 22 cycles~day$^{-1}$ were found to have a much lower
significance in the {\sc clean}ed analysis and were thus discounted as
an artifact of the windowing. Both remaining peaks above the 4$\sigma$ level
(1\,842.4$\pm$1.5~s and 2\,645.0$\pm$4.0~s) were equally viable
periods. We selected the 1\,842.4~s period to fold the data at, but
we stress that the other period was an equally likely candidate
period, see Fig.~\ref{J17195_2_10kev_folded}. Folding the data in
each energy band at this period showed that the modulation depth is
constant across them all (see Table~\ref{modulation_depths}). We
also note that there is a further peak (at just below $3\sigma$
significance) at 941~s, whose period is close to half that of the
1842.4~s candidate period, and may therefore represent a first
harmonic.

\begin{figure}
  \resizebox{0.8\hsize}{!}{\rotatebox{-90}{\includegraphics{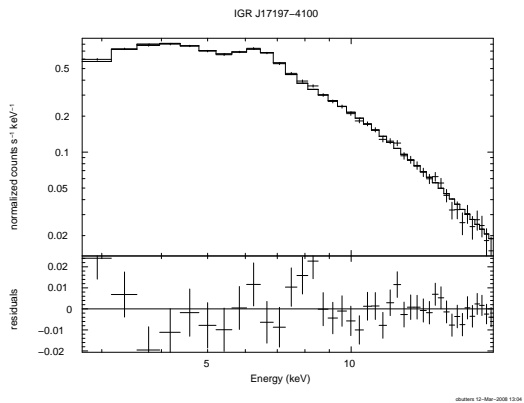}}}
  \caption{2.5--20~keV mean spectrum of J1719 fitted with a photoelectrically
  absorbed power law plus iron line profile.}
  \label{J17195_spectrum}
\end{figure}

Spectral analysis showed the presence of an iron line in a
photoelectrically absorbed bremsstrahlung profile, however the fit
was poor with $\chi^2_{\rm reduced}=3.0$. A better fit was
achieved with a power law model as shown in
Fig.~\ref{J17195_spectrum} and Table~\ref{spectral_fits}, however
this fit had the column density pegged to a lower limit of
0.7~$\times10^{22}$~cm$^{-2}$ to reflect the galactic column
density.

\subsection{XSS~J12270--4859}
Data were collected over the course of just over one day (see
Table~\ref{observing_log}). Total good time on target (26\,814~s)
was split over nine segments. The raw target count rate varied between
2.4--10.9~ct~s$^{-1}$~PCU$^{-1}$, the generated background count
rate varied between 2.8--3.9~ct~s$^{-1}$~PCU$^{-1}$.

\begin{figure}
  \resizebox{0.8\hsize}{!}{\rotatebox{90}{\includegraphics{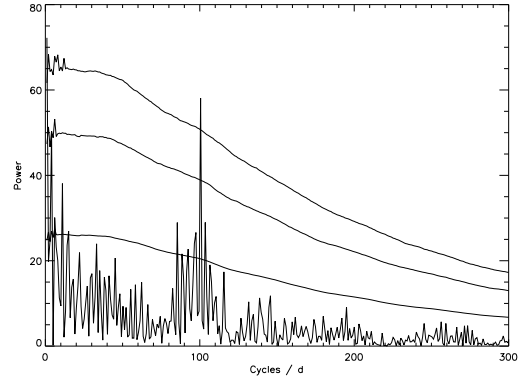}}}
  \caption{2--10~keV periodogram of J1227. Three significance
  levels, 95.2, 99.7 and 99.954\% (2, 3 and 4$\sigma$ respectively), are superimposed.}
  \label{J12270_red_noise}
\end{figure}

\begin{figure}
  \resizebox{0.8\hsize}{!}{\includegraphics{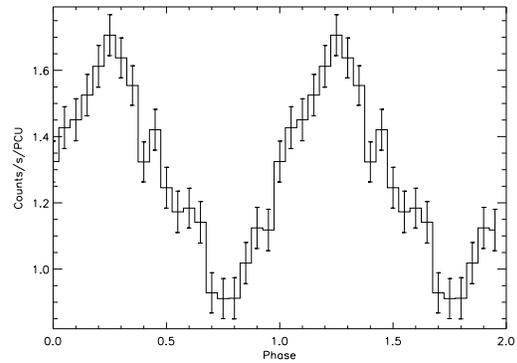}}
  \caption{2--10~keV lightcurve of J1227 folded at the 859.57~s pulse period
  with an arbitrary zero point. Two cycles are shown for clarity.}
  \label{J12270_folded}
\end{figure}

Analysis of the lightcurve showed significant ($>4\sigma$) structure
at $\sim100$~cycles~day$^{-1}$ (see Fig.\ref{J12270_red_noise}). The
peak of this structure was at 859.57$\pm$0.64~s. Folding the data at this
period showed a clear modulation  in the 2--10~keV energy band (see
Fig.~\ref{J12270_folded}), with approximately the same
percentage depth in each energy band (see
Table~\ref{modulation_depths}). There was also a peak at approximately
one~cycle day$^{-1}$ in the periodogram; we discounted this peak as it was
of the order of the length of the observing run, and was probably
a feature of the window function.

\begin{figure}
  \resizebox{0.8\hsize}{!}{\rotatebox{-90}{\includegraphics{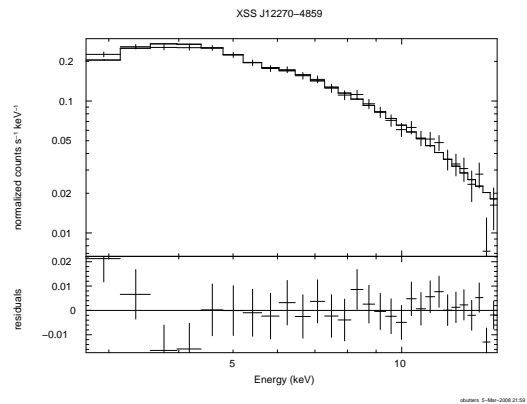}}}
  \caption{2--15~keV mean spectrum of J1227 fitted with a photoelectrically
  absorbed power law plus Gaussian.}
  \label{J12270_spectrum}
\end{figure}

In fitting the spectrum, the column density was again pegged to the
lower limit of the galactic column density for both the models. The best
fit was the power law model, giving $\chi^2_{\rm reduced}=0.8$
(see Table~\ref{spectral_fits}). There is no significant sign of
an excess at the iron line energy (see Fig.\ref{J12270_spectrum}).
A Gaussian was fitted to the expected position of the iron emission feature,
but in each case only a small upper limit to the equivalent width was
found ($<0.08$ and $<0.17$~keV for models A and B respectively).

\section{Discussion}

\subsection{XSS J00564+4548}

We interpret the period found here (465.68$\pm$0.07~s) as the
spin period of the WD in J0056. \citet{bikmaev06} gave an approximate
value of 480~s from their analysis. To obtain an
estimate of the error in their period we consider the FWHM of their
Lomb-Scargle plot,  which gives 480$\pm$20~s. Our period
determination therefore is in agreement with their optical data.
If we interpret the second strongest peak in our periodogram as
a beat period, this implies the orbital period must be $\sim2.7$~hrs.
This places it in the 2--3~hr CV period gap, but we note that there
are now several CVs in this range and some mCVs too. This interpretation
would also explain the cluster of peaks around the spin period as being
various harmonics of the beat period. We are unsure of the origin of the
41~cycles~day$^{-1}$ peak as it is too short to be interpreted as an
orbital period of a typical IP.
The energy dependant modulation depth of the folded lightcurves is common among
IPs and indicates an accretion column absorbing structure \citep{norton89}.
We also note that the pulse profile of J0056 is very similar in shape to that of
\object{FO~Aqr} \citep{beardmore98}.

The spectral fits indicate that the emission process is more likely to
be a photoelectrically absorbed single temperature bremsstrahlung
process rather than a power
law, as is common in IPs. The power law spectral fit is however in
agreement with \citet{revnivtsev04}. The Gaussian at 6.5~keV is identifiable
with an iron feature which is also a common aspect of IPs, and agrees with
the feature found by \citet{bikmaev06}.

The {\em ROSAT} bright source catalogue has no other sources in
the {\em RXTE} PCA field of view of J0056. \cite{bikmaev06} showed
that there was a source in the error circle of {\em ROSAT} but it was
only present at low energies, and so will not affect this measurement. There is
therefore very little contamination from other near by sources.
We note that J0056 is significantly brighter now than was reported in
the {\em RXTE} all sky survey \citep{revnivtsev04}.

\subsection{IGR J17195--4100}
The peaks in the periodogram of J1719 (1\,842.4$\pm$1.5~s and
2\,645.0$\pm$4.0~s) are typical for a spin period length in IPs. The small
peak at 941~s is close to being half the 1\,842.4 period, and
therefore may be a first harmonic, however, the significance of
this peak is below $3\sigma$. We note that if the two longer periods
above correspond to the spin and beat periods respectively then
this implies an orbital period of approximately 1.7~hrs.
The small, almost constant, modulation depth seen in the lightcurves in
each energy band is not present among any other confirmed IPs and
implies that the modulation is caused by obscuration as opposed to absorption.

The presence of an iron feature at 6.5~keV is a strong indicator
of an IP classification. A significantly better spectral
fit is obtained from a power
law instead of a bremsstrahlung model. The 2--10~keV fluxes obtained
from each of the spectral models are also considerably larger than the
value reported by \citet{tomsick06} in the 0.3--10~keV energy
band. This may be indicative of the simplistic single temperature
  bremsstrahlung model used here; multi-temperature fits are often
  needed to model the post-shock flow (see
  e.g. \citet{ezuka99}). However, the signal to noise and the spectral
  resolution of the data is such that a complex model may yield non-unique or degenerate results.

There are several X-ray sources near by in the PCA field of view which may
contribute to the count rate. We used the {\em ROSAT} count rate of
each source to estimate an {\em RXTE} count rate using the on line tool
webPIMMS\footnote{http://heasarc.nasa.gov/Tools/w3pimms.html} (for each source
we assume a power law with a photon index of 1.7), we then scaled this value
by the response of the PCAs based upon the distance from the source. For this
source the contribution is up to 1.1~ct~s$^{-1}$~PCU$^{-1}$ in
the 2--20~keV energy band, i.e. $\lesssim20\%$ of the measured count
rate. These extra sources will have the effect of decreasing the
percentage modulation depth. Moreover, since it is likely that
  these sources are softer than the target, the contamination will
  have a greater effect at lower energies. The modulation
  depth will therefore be reduced more at lower energies. This could
  make a decreasing modulation depth with energy look like a constant
  modulation depth with energy. The spectral fitting is likely to be
affected by these other sources and they may be the cause
of the poor bremsstrahlung model fit. It is also possible these other
sources may skew the model fit in such a way that the calculated flux
is then overestimated, this may explain why the flux reported here is
larger than the \citet{tomsick06} value. We emphasise that this is only an
estimate of the contamination; the other sources may differ markedly from
the assumed spectral shape.

\subsection{XSS J12270--4859}
J1227 exhibits a structure in the periodogram that indicates it has a
period close to 100~cycles~day$^{-1}$. This is consistent with being
interpreted as a spin period. It shows an approximately constant
modulation in each energy band at the 859.57$\pm$0.64~s period, which
implies that the process causing this effect must be
a geometrical effect causing obscuration instead of absorption.

The upper limit placed on the equivalent width of a potential iron
line is small, and goes against the classification of this as an
IP, since all IPs exhibit some kind of iron emission features. We
  do note however that \citet{masetti06} did see significant iron
  features in their optical spectra. The best spectral fit is obtained
from a power law profile, the parameters of which are in good agreement with
\citet{revnivtsev04}. Again we note that a multi-temperature
  bremsstrahlung fit may be more accurate, but beyond the scope of this
  study. The count rate has not changed significantly since the measurements of
\citet{revnivtsev04}.

This source also has nearby X-ray sources that may contribute to the count rate.
Using the same procedure as outlined above we estimate that
they may have contributed up to 0.26~ct~s$^{-1}$~PCU$^{-1}$ in the
2--20~keV energy band, i.e $\lesssim20\%$ of the total
count rate. It is again possible that these extra sources would
  alter the modulation depths, and that the spectral fits are also skewed.

\section{Conclusion}
The unambiguous detection of an X-ray spin period of 465.68$\pm$0.07~s
in J0056 and its decreasing modulation depth with increasing energy,
along with its spectral properties, confirm its inclusion into
the IP class. Both J1719 and J1227 clearly exhibit some
properties seen in IPs, but not to an extent for us to
definitively classify them as such. We do note that it is likely these
latter two are IPs, and that their true nature is being masked by the presence of
contamination from other sources. X-ray imaging of these sources will definitively
decide their fate, allowing their true spectral characteristics to be
revealed. All three sources would benefit from long base
line optical campaigns to determine their orbital periods and
ratify the validity of the periods in J1719 and J1227.
If J1227 does turn out to be an IP, then the presence of an X-ray iron
feature will have to be reconsidered as a defining characteristic of
IPs, since it is not present here.

\bibliographystyle{aa}
\bibliography{9942}

\begin{thebibliography}{18}
\expandafter\ifx\csname natexlab\endcsname\relax\def\natexlab#1{#1}\fi

\bibitem[{{Barlow} {et~al.}(2006){Barlow}, {Knigge}, {Bird}, {J Dean}, {Clark},
  {Hill}, {Molina}, \& {Sguera}}]{barlow06}
{Barlow}, E.~J., {Knigge}, C., {Bird}, A.~J., {et~al.} 2006, MNRAS, 372, 224

\bibitem[{{Beardmore} {et~al.}(1998){Beardmore}, {Mukai}, {Norton}, {Osborne},
  \& {Hellier}}]{beardmore98}
{Beardmore}, A.~P., {Mukai}, K., {Norton}, A.~J., {Osborne}, J.~P., \&
  {Hellier}, C. 1998, MNRAS, 297, 337

\bibitem[{{Bikmaev} {et~al.}(2006){Bikmaev}, {Revnivtsev}, {Burenin}, \&
  {Sunyaev}}]{bikmaev06}
{Bikmaev}, I.~F., {Revnivtsev}, M.~G., {Burenin}, R.~A., \& {Sunyaev}, R.~A.
  2006, Astronomy Letters, 32, 588

\bibitem[{{Bird} {et~al.}(2004){Bird}, {Barlow}, {Bassani}, {Bazzano},
  {Bodaghee}, {Capitanio}, {Cocchi}, {Del Santo}, {Dean}, {Hill}, {Lebrun},
  {Malaguti}, {Malizia}, {Much}, {Shaw}, {Stephen}, {Terrier}, {Ubertini}, \&
  {Walter}}]{bird04}
{Bird}, A.~J., {Barlow}, E.~J., {Bassani}, L., {et~al.} 2004, ApJL, 607, L33

\bibitem[{{Bird} {et~al.}(2007){Bird}, {Malizia}, {Bazzano}, {Barlow},
  {Bassani}, {Hill}, {B{\'e}langer}, {Capitanio}, {Clark}, {Dean}, {Fiocchi},
  {G{\"o}tz}, {Lebrun}, {Molina}, {Produit}, {Renaud}, {Sguera}, {Stephen},
  {Terrier}, {Ubertini}, {Walter}, {Winkler}, \& {Zurita}}]{bird07}
{Bird}, A.~J., {Malizia}, A., {Bazzano}, A., {et~al.} 2007, ApJS, 170, 175

\bibitem[{{Bradt} {et~al.}(1993){Bradt}, {Rothschild}, \& {Swank}}]{bradt93}
{Bradt}, H.~V., {Rothschild}, R.~E., \& {Swank}, J.~H. 1993, A\&AS, 97, 355

\bibitem[{{Butters} {et~al.}(2007){Butters}, {Barlow}, {Norton}, \&
  {Mukai}}]{butters07}
{Butters}, O.~W., {Barlow}, E.~J., {Norton}, A.~J., \& {Mukai}, K. 2007, A\&A,
  475, L29

\bibitem[{{Dickey} \& {Lockman}(1990)}]{dickey90}
{Dickey}, J.~M. \& {Lockman}, F.~J. 1990, ARAA, 28, 215

\bibitem[{{Ezuka} \& {Ishida}(1999)}]{ezuka99}
{Ezuka}, H. \& {Ishida}, M. 1999, ApJS, 120, 277

\bibitem[{{Hakala} {et~al.}(2004){Hakala}, {Ramsay}, {Wheatley}, {Harlaftis},
  \& {Papadimitriou}}]{hakala04}
{Hakala}, P., {Ramsay}, G., {Wheatley}, P., {Harlaftis}, E.~T., \&
  {Papadimitriou}, C. 2004, A\&A, 420, 273

\bibitem[{{Lehto}(1997)}]{lehto97}
{Lehto}, H.~J. 1997, in Applications of time series analysis in astronomy and
  meteorology, ed. T.~{Subba Rao}, M.~B. {Priestley}, \& O.~{Lessi} (London
  Chapman and Hall)

\bibitem[{{Masetti} {et~al.}(2006){Masetti}, {Morelli}, {Palazzi}, {Galaz},
  {Bassani}, {Bazzano}, {Bird}, {Dean}, {Israel}, {Landi}, {Malizia},
  {Minniti}, {Schiavone}, {Stephen}, {Ubertini}, \& {Walter}}]{masetti06}
{Masetti}, N., {Morelli}, L., {Palazzi}, E., {et~al.} 2006, A\&A, 459, 21

\bibitem[{{Norton} \& {Watson}(1989)}]{norton89}
{Norton}, A.~J. \& {Watson}, M.~G. 1989, MNRAS, 237, 853

\bibitem[{{Pandey} {et~al.}(2006){Pandey}, {Rao}, {Manchanda}, {Durouchoux}, \&
  {Ishwara-Chandra}}]{pandey06}
{Pandey}, M., {Rao}, A.~P., {Manchanda}, R., {Durouchoux}, P., \&
  {Ishwara-Chandra}, C.~H. 2006, A\&A, 453, 83

\bibitem[{{Ramsay} {et~al.}(2008){Ramsay}, {Wheatley}, {Norton}, {Hakala}, \&
  {Baskill}}]{ramsay08}
{Ramsay}, G., {Wheatley}, P.~J., {Norton}, A.~J., {Hakala}, P., \& {Baskill},
  D. 2008, ArXiv e-prints, 804

\bibitem[{{Revnivtsev} {et~al.}(2004){Revnivtsev}, {Sazonov}, {Jahoda}, \&
  {Gilfanov}}]{revnivtsev04}
{Revnivtsev}, M., {Sazonov}, S., {Jahoda}, K., \& {Gilfanov}, M. 2004, A\&A,
  418, 927

\bibitem[{{Tomsick} {et~al.}(2006){Tomsick}, {Chaty}, {Rodriguez}, {Foschini},
  {Walter}, \& {Kaaret}}]{tomsick06}
{Tomsick}, J.~A., {Chaty}, S., {Rodriguez}, J., {et~al.} 2006, ApJ, 647, 1309

\bibitem[{{Warner}(1995)}]{warner95}
{Warner}, B. 1995, {Cataclysmic variable stars} (Cambridge Astrophysics Series,
  Cambridge, New York: Cambridge University Press, |c1995)

\end{thebibliography}

\end{document}